\documentclass[aps,prl,twocolumn,showpacs,showkey,square,numbers,amssymb,amsmath,nobibnotes]{revtex4-1}
\usepackage{bm}
\usepackage{times,float}
\usepackage{graphicx}
\usepackage[usenames,dvipsnames,svgnames]{xcolor}
\usepackage{hyperref}
\hypersetup{colorlinks=true, linkcolor=NavyBlue, citecolor=PineGreen,urlcolor=cyan}

\newcommand{\ket}[1]{\left|#1\right\rangle}
\newcommand{\bra}[1]{\left\langle #1\right|}
\newcommand{\bracket}[1]{\left\langle #1\right\rangle}

\newcommand{\beeq}[1] {\begin{equation}\begin{split}#1\end{split}\end{equation}}

\newcommand{\Eq}[1]{Eq.\,\eqref{#1}}

\newcommand{\Fig}[1]{Fig.\,\ref{#1}}

\begin{document}

\title{The exact two-spinon longitudinal dynamical structure factor of the anisotropic XXZ model}

\author{Isaac P\'erez Castillo}
\address{Department of Quantum Physics and Photonics, Institute of Physics, UNAM, P.O. Box 20-364, 01000 Mexico City, Mexico}
\address{London Mathematical Laboratory, 18 Margravine Gardens, London W6 8RH, United Kingdom}
 
\begin{abstract}
Inelastic neutron scattering experiments are commonly used to unveil how  excitations on Heisenberg spin models play a role in dynamical correlations functions.  For a certain class of materials, like CsCoCl$_3$ or CsCoBr$_3$ salts, it turns out that their  magnetic properties are  fairly well approximated by quasi-one dimensional XXZ models, which enjoy the property of quantum integrability. In these instances, one can in principle use their underlying algebraic structure to  describe very precisely how excitations, the so-called spinons, participate in dynamical correlations functions.  Even though the available theories (either algebraic Bethe ansatz or quantum group approach)  provide all the needed physical quantities such as form factors, complete set of eigenstates and spectrum,  it is typically a rather daunting task, however, to obtain sufficiently simple analytical expressions for computing  Dynamical Structure Factors (DSFs), valuable, e.g.,  for parameters estimation  based on experimental data.  This is particularly the case for the longitudinal DSF of the XXZ model, which has eluded a formal mathematical treatment thus far.  Using the quantum group approach, we present here an exact and simple expression of the 2-spinon longitudinal DSF and show  our results to be consistent with the expected sum rules and the isotropic and Ising antiferromagnet limiting cases. 
\end{abstract}

\maketitle

Heisenberg spin chains \cite{Heisenberg1928}  represent a long-standing arena to introduce,  test, and deeply understand seminal concepts in strongly correlated quantum systems. While the eigenstates and eigenvectors of the one-dimensional quantum spin chain  with spin $S=1/2$  have been known ever since  Hans Bethe's original work \cite{Bethe1931}, it took  around 60 years to get a handle on  the properties of its ground state and  its excitations, the so-called spinons \cite{Faddeev1981}.  It turned out along the way that spin $S=1/2$ Heisenberg chains with nearest neighbor interactions and other similar models have the very attractive property  of quantum integrability \cite{Korepin1993,Takahashi1999}. This allows one to understand and, in principle, to control very precisely the nature of these excitations and their impact in dynamical correlation functions.  While all the mathematical ingredients to obtain  dynamical correlation functions such as  eigenstates, eigenvalues and form factors, etc.  are readily available, it turned out that correlations involving the $z$ component of the spin operator posed a very difficult task.  This situation has  been ignored for a while due to the fact that inelastic neutron scattering measurements of experimentally available quasi-1D Heisenberg antiferromagnets,  as for instance the Ising-type materials CsCoCl$_3$ \cite{yoshizawa1981,goff1995} and CsCoBr$_3$ \cite{nagler1982,nagler1983}, do not require the knowledge of the longitudinal DSF of the spin operator.  However,  recent experiments performed on Yb$_2$Pt$_2$Pb do need those formulas since, due to a strong anisotropy of the Land\'e $g$-factor, only the longitudinal correlation can be measured by neutron scattering \cite{wu2016}.

The main goal of the present Letter is to present  an exact and compact expression for the two-spinon contribution to the longitudinal  DSF using quantum group approach and assess the correctness of  our analytical findings with some expected sum rules. Our formulas are very simple and compact and may be used to draw some conclusions on whether  emergent Hamiltonians for newly studied materials, such as  Yb$_2$Pt$_2$Pb, are correctly captured by the XXZ model with only nearest-neighbour interactions.

As we will be using the results from the quantum group approach \cite{Jimbo1995}, we will  directly tackle the  spin $S=1/2$ XXZ antiferromagnetic chain of infinite length  given by the  Hamiltonian:
\begin{equation}
      H_{\text{XXZ}} = -J \sum_{n=-\infty}^{\infty}
      \left(S^x_{n}S_{n+1}^x+S^y_{n}S_{n+1}^y+\Delta S^z_{n}S_{n+1}^z\right)\,,
\end{equation}
where $S^{x,y,z}_{n}$ are the spin-1/2 operators acting on site $n$, and $\Delta$ is the anisotropy parameter. We will focus on the massive regime with   $-\infty<\Delta<-1$.  The limit $\Delta\to-\infty$ corresponds to the Ising antiferromagnet,  around which, most of its  properties can be easily calculated using perturbation theory. In particular, the excitations above the doubly degenerate ground states (N\'eel states) correspond to domain walls \cite{villain1975}, i.e., spinons, which can be envisaged as solitons of unit length in the lattice space. Moreover, in the vicinity of the Ising antiferromagnetic  point, domain-wall pair states are the  excitations that mainly contribute to   neutron scattering amplitudes \cite{ishimura1980,nagler1982}. 

Within the algebraic approach of the celebrated Kyoto school \cite{Jimbo1995}, it is known that multi-spinon exctations, denoted as $\ket{\{\xi\}_m}_{\{\epsilon\}_m;(i)}$ with $\{\xi\}_m=\{\xi_1,\ldots,\xi_m\}$ and $\{\epsilon\}_m=\{\epsilon_1,\ldots,\epsilon_m\}$, can be generated starting from the doubly degenerate ground state $\ket{\text{vac}}_{(i)}$ for given $\Delta<-1$ with $i=0,1$.
Each spinon indexed by $j=1,\ldots,m$ is characterised by a pair $(\xi_j,\epsilon_j)$, where the spectral parameter $\xi_j\in\{\mathbb{C}: |\xi_j|=1\}$ lives on the complex unit circle while $\epsilon_j\in\{-,+\}$ labels the spinon's spin orientation. The $m$ spinon exctations are exact eigenstates of the Hamiltonian, and as such also translationally invariant,
\beeq{
   H_{\text{XXZ}}\ket{\{\xi\}_m}_{\{\epsilon\}_m;(i)}
     &= E(\{\xi\}_m)\ket{\{\xi\}_m}_{\{\epsilon\}_m;(i)}\,,
\\
   T \ket{\{\xi\}_m,}_{\{\epsilon\}_m;(i)}
     &=e^{iP(\{\xi\}_m)}\ket{\{\xi\}_m,}_{\{\epsilon\}_m;(1-i)}
\text{ .}
}
Here $T$ denotes the translation operator by one lattice site,  $E(\{\xi\}_m){=}\sum_{j=1}^m e(\xi_j)$ and $P(\{\xi\}_m){=}\sum_{j=1}^m p(\xi_j)$. Henceforth, we will use an elliptic parametrization to represent spinons. We choose $\xi{=}ie^{i\pi\beta/2K}$ with $-K\leq \beta< K$, where $K{\equiv} K(k)$ is the complete elliptic integral of the first kind. The anisotropy parameter becomes $\Delta=-\cosh\bigl(\frac{\pi K'}{K}\bigr)$, where  $K'{\equiv} K(k')$, and $k'{\equiv}\sqrt{1-k^2}$ begin  the complementary elliptic modulus. The spinon's energy and momentun take the simple form:
\beeq{
   e(\beta) &=
   I\text{dn}(\beta)\,,\quad p(\beta)=\text{am}(\beta)+\tfrac{\pi}{2}
\label{eq:edisp}
}
with $\text{dn}(x)\equiv\text{dn}(x,k)$ and $\text{am}(x)\equiv\text{am}(x,k)$ the usual Jacobi elliptic functions with elliptic modulus $k$ and $I\equiv \frac{J K}{\pi}\sinh\bigl(\frac{\pi K'}{K}\bigr)$ is the energy scale of the spinon.

The role that  spinons play in dynamical spin-spin correlations functions can be measured  experimentally by  neutron scattering. We focus on the longitudinal DSF at zero temperature, which takes the following form for the infinite chain:
\beeq{
   S^{zz}(Q,\omega) &= 
   \sum_{n=-\infty}^{\infty} \int_{-\infty}^\infty
   dt \, e^{i(\omega t-Qn)}\bracket{S_n^z(t)S_0^{z}(0)}\,,
\label{eq:DSF}
}
where the bracket $\bracket{(\cdots)}$ corresponds to averaging with respect to the two ground state vacua \footnote{At present, finite temperature calculations using quantum group approach are  out of reach.}.  Using the results of quantum group approach, $S^{zz}(Q,\omega)$ can be rewritten  as a series Lehmann representation that emphasizes the physical role that the multispinon excitations play into  this two-point correlation function:
\beeq{
   S^{zz}(Q,\omega)&=\sum_{m \text{ even }\geq 0}S^{zz}_{(m)}(Q,\omega)\,,
}
where $S^{zz}_{(m)}(Q,\omega)$ is the contribution from transitions between the two ground states and the m-spinon states. These transitions naturally involved the form factors $\,_{(i)}\bra{\text{vac}}S^{z}_{0} \ket{\{\xi\}_m}_{\{\epsilon\}_m;(i)}$ connecting the two vacua with the m-spinon states via the $z$ component of the spin operator. From the whole sum in Eq. \eqref{eq:DSF}, we will focus on the contributions $m=0$ and $m=2$, as they carry most of the weight of $S^{zz}(Q,\omega)$ for a wide range of values of the anisotropy parameter $\Delta$, as we will see below by inspecting several sum rules.

The zeroth contribution $S^{zz}_{(0)}(Q,\omega)$ can be easily derived \footnote{See Supplemental Information for details.}.  Unlike the  transverse case, it is non-zero  and directly related  to the squared staggered {\it static} background magnetization, thus contributing at $\omega=0$ only:
\beeq{
   S^{zz}_{(0)}(Q,\omega) = \pi^2\left( \tfrac{(q^2_e;q^2_e)_{\infty}}{(-q^2_e;q^2_e)_{\infty}}
   \right)^4
   \ \delta_{Q,0} \, \delta(\omega)
\label{eq:LDSF0}
}
where $q_e=e^{-\frac{\pi K'}{K}}$ is the so-called elliptic nome and the notation $(a;q)_\infty$ corresponds to the $q$-Pochhammer symbol.

\begin{figure*}[tb]
\begin{center}
\includegraphics[height=5.cm,width=8.25cm]{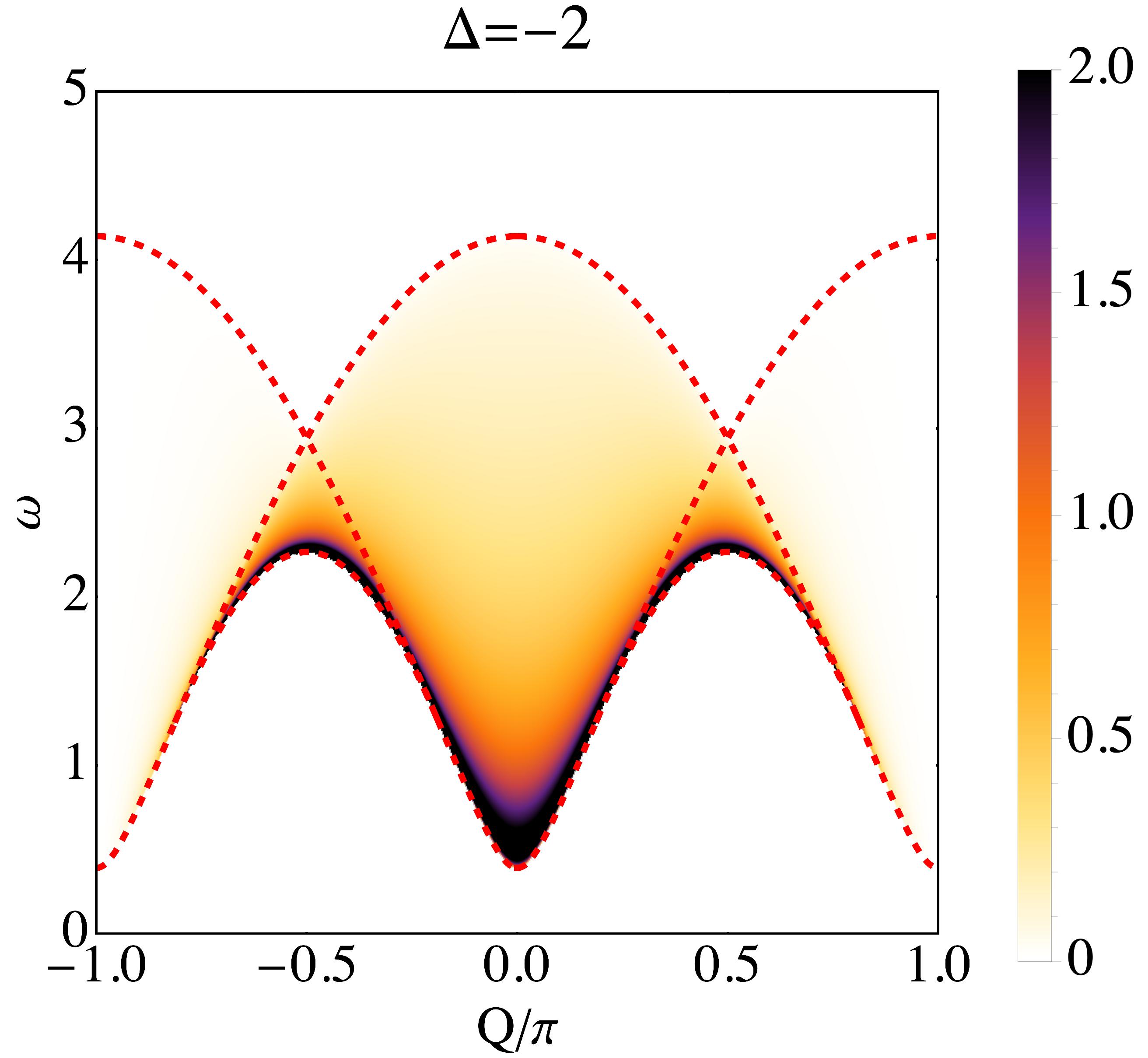} 
\includegraphics[height=5.cm,width=8.25cm]{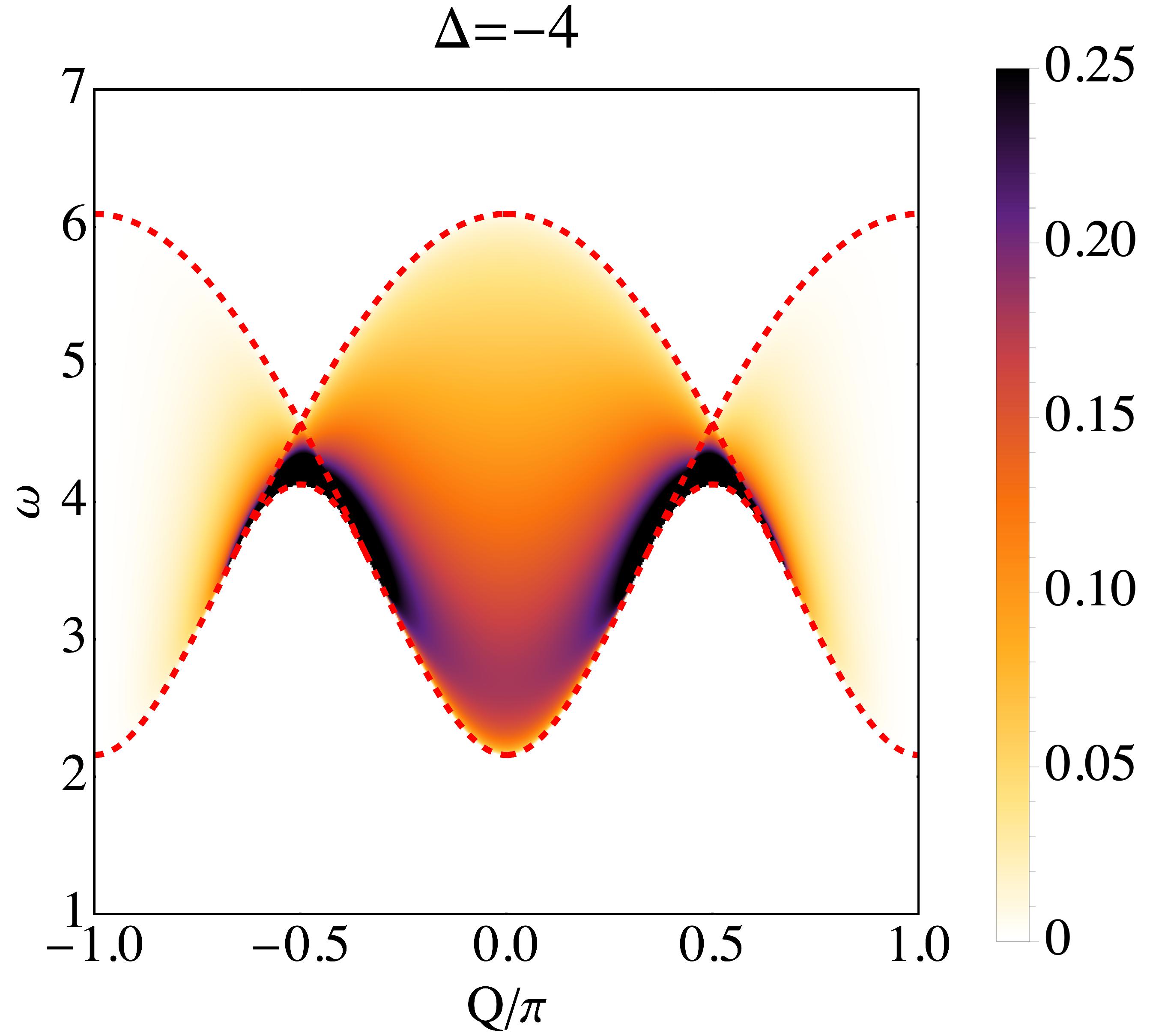}\\
\includegraphics[height=5.cm,width=8.25cm]{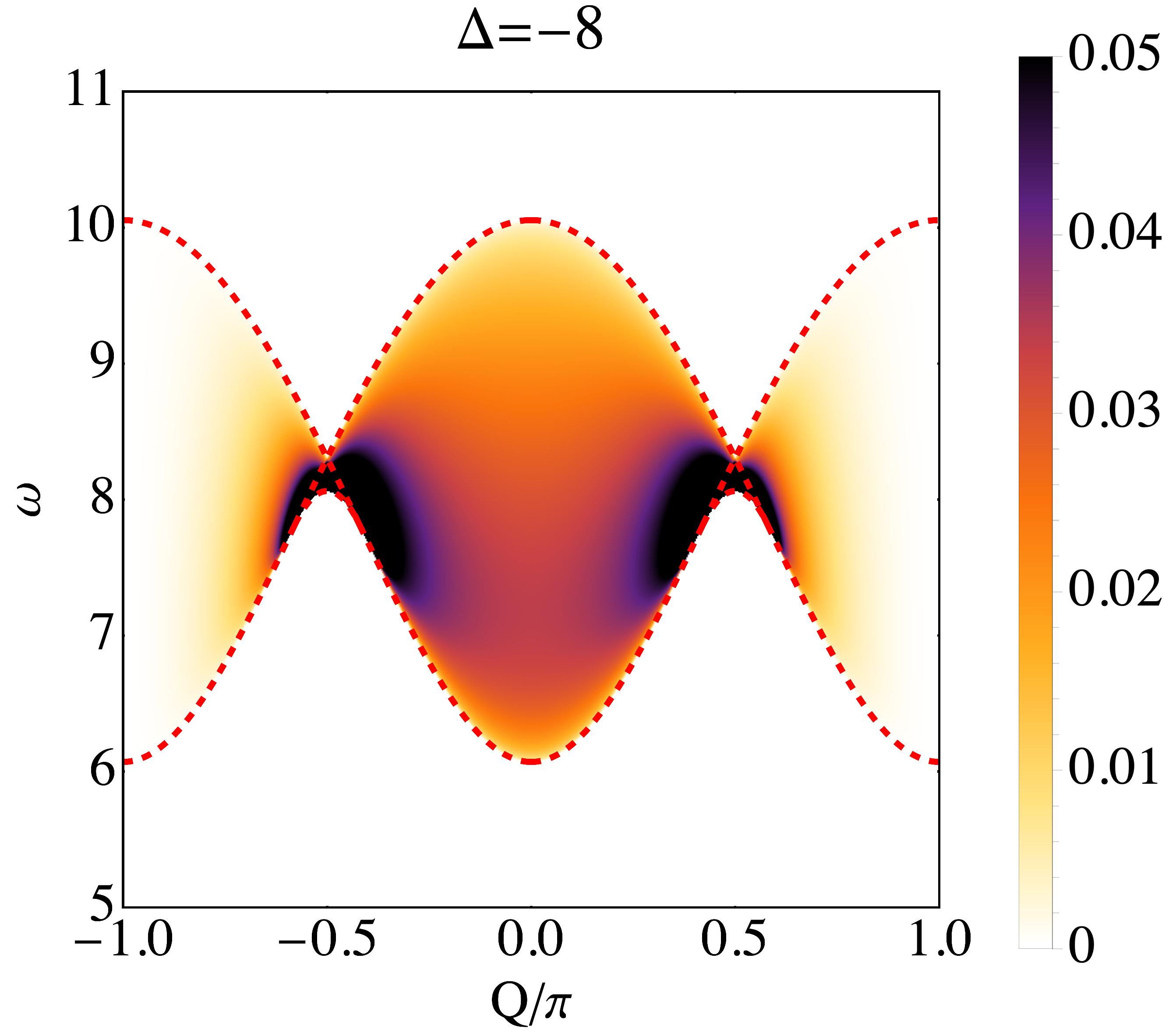} 
\includegraphics[height=5.cm,width=8.25cm]{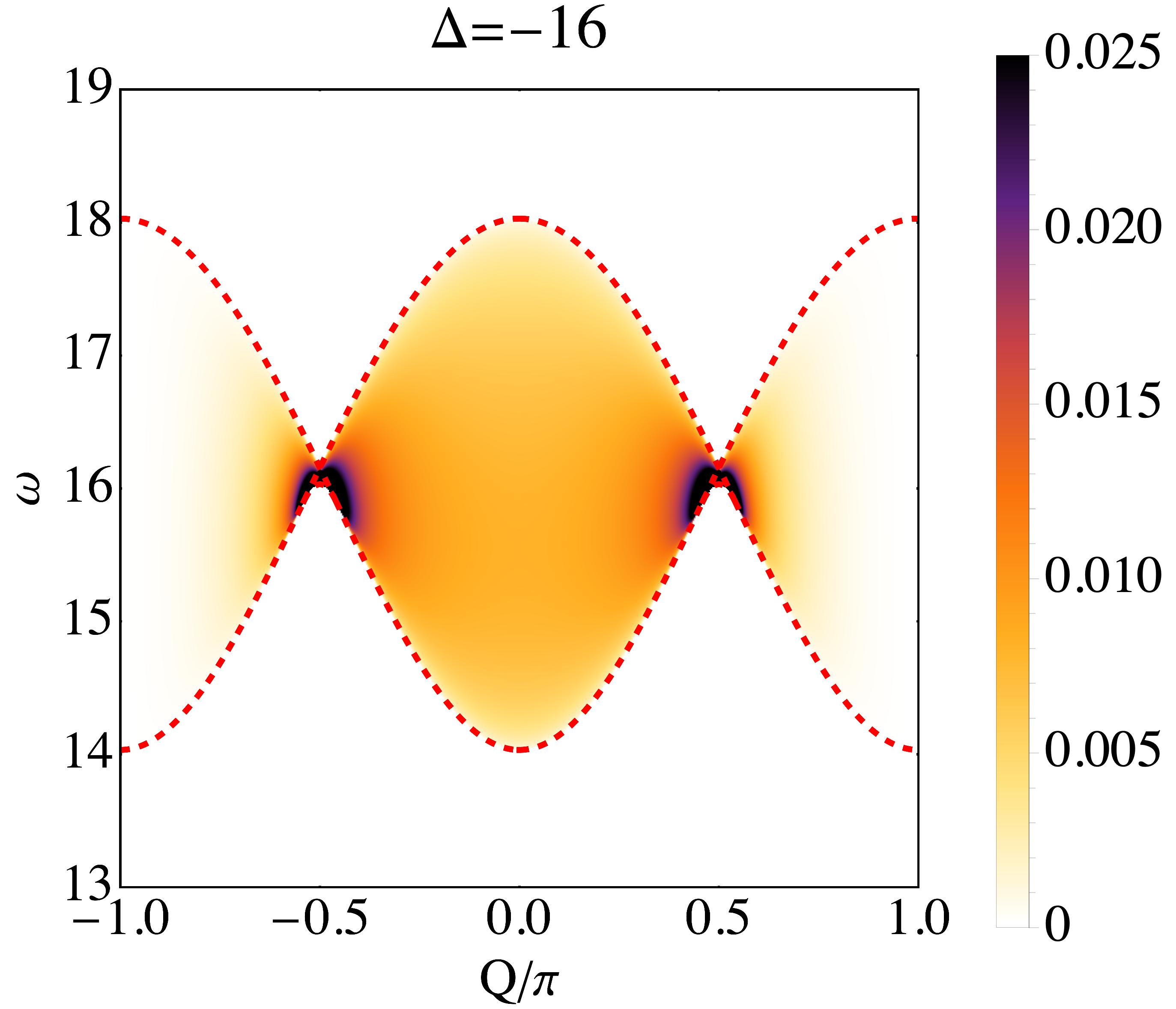}
\caption{Density plot  of $S^{zz}_{(2)}(Q,\omega)$ in the $(Q,\omega)$-plane for anisotropy values of $\Delta=-2$, $-4$, $-8$, and $\Delta=-16$.  Red dashed lines mark the analytical boundaries of  support of the DSF.}
\label{fig:1}
\end{center}
\end{figure*} 
The next  contribution, $S^{zz}_{(2)}(Q,\omega)$, comes from non-trivial  transitions from the two ground states to the two-spinon continuum band. Mathematical progress in the evaluation of this term has been hindered due to the inability of evaluating an essential singularity \cite{Isaac2016} of the two-spinon form factor formula provided by \cite{Jimbo1995}. Luckily, there are recent alternatives to obtain this form factor \cite{Lashkevich2002,Dugave2015,Dugave2015b,Isaac2016}. Thus, resolving the two-spinon continuum dispersion relation correctly \cite{Perez2008}, and after lengthy and tedious mathematical manipulations \cite{Note1}, we are able to obtain the following simple exact expression of the two-spinon longitudinal DSF:
\begin{eqnarray}
&& S^{zz}_{(2)}(Q,\omega) =
   \sqrt{q_e} k \, \tfrac{\omega^2+\kappa \omega_0^2+B}{\omega^3 B}
   \sum_{\sigma=\pm1} 
   \tfrac{1+\sigma \cos(Q)}{ W_\sigma} \,\times
\label{eq:LDSF2} \\
 && \ 
   \tfrac{\vartheta^2_A(\beta^{(\sigma)}_{-})}{\vartheta^2_d(\beta^{(\sigma)}_{-})}
  \tfrac{\omega^2 -\sigma (B- \kappa\omega_0^2)}%
        {|\Delta| -\sigma \cos\bigl(\frac{\pi\beta^{(\sigma)}_{-}}{K}\bigr)}
   \Bigl(
      \tfrac{1-\kappa}{1+\kappa} \, \delta_{\sigma,+}
    + \delta_{\sigma,-}
   \Bigr)
 \, \mathbb{I}_{(Q,\omega)\in C_\sigma(Q,\omega)}
\,. \notag
\end{eqnarray}
The expression provided by \Eq{eq:LDSF2} is the main result of the present Letter.  Here, the support of the DSF is given by the indicator function $\mathbb{I}_{(Q,\omega)\in C_\sigma(Q,\omega)}$, equal to 1  if the point $(Q,\omega)$ lies within the continuum sheet $C_\sigma(Q,\omega)$, and 0  otherwise. These two sheets for $\sigma\in\{-1,+1\}$ result from the  overlapping of a two-spinon dispersion relation band with another one shifted by $\pi$, everything modulus $2\pi$. As a result, the lower and upper boundaries of the sheet $C_{+}(Q,\omega)$ are given by \cite{bougourzi1998,Perez2008}:
\beeq{
   \Omega_{\text{lo}}(Q)=\left\{
   \begin{array}{ll}
      \omega_0(Q)&Q\in[Q_{\kappa},\pi-Q_{\kappa}]\\
      \omega_+(Q)&Q\in[\pi-Q_{\kappa},\pi]\\
   \end{array}
   \right.\,,
}
and
\beeq{
   \Omega_{\text{up}}(Q)=\omega_-(Q)\,, \qquad Q\in[Q_\kappa,\pi]\,,
}
respectively, with the definitions [using $I$ as with \Eq{eq:edisp}]:
\beeq{
   \kappa&=\tfrac{1-k'}{1+k'}\,,\quad Q_\kappa=\mathrm{acos}(\kappa)\,,\\
   \omega_{\pm}(Q)&=\tfrac{2I}{1+\kappa}\sqrt{1+\kappa^2\pm2\kappa\cos(Q)}\,,\\
   \omega_{0}(Q)&=\tfrac{2I}{1+\kappa}\sin(Q)\,.
}
The continuum sheet $C_{-}(Q,\omega)$ is simply the sheet $C_{+}(Q,\omega)$ reflected around $\pi/2$. The expression for $\beta_{-}^{(\sigma)}$  in \Eq{eq:LDSF2} comes from solving the two-spinon dispersion relation and takes the following form
\cite{Perez2008,bougourzi1998,Note1}:
\beeq{
   \beta_{-}^{(\sigma)}(Q,\omega) &= \text{dn}^{-1}
   \Bigl(
      \tfrac{1+\sigma\cos(Q)}{|\sin(Q)|}
      \sqrt{\tfrac{\omega^2-\kappa\omega_0^2(Q)+B}{\omega^2+\kappa\omega_0^2(Q)-B}},k
   \Bigr)\,,
}
while the functions $B\equiv B(Q,\omega)$ and $W_\sigma \equiv W_{\sigma}(Q,\omega)$ also appearing in Eq. \eqref{eq:LDSF2} read
\beeq{
   B(Q,\omega) &=
      \sqrt{\omega^2-\kappa^2\omega_0^2(Q)}
      \sqrt{\omega^2-\omega_0^2(Q)}\,
\\
   W_\sigma(Q,\omega) &= \sqrt{
      \kappa^2\tfrac{\omega_0^4(Q)}{\omega^4} -
      \left(\tfrac{B(Q,\omega)}{\omega^2}+\sigma\cos(Q)\right)^2
   }\,,
}
respectively. Finally, $\vartheta_d(\beta^{(\sigma)}_{-})$ refers to Neville's theta function while the function
$\vartheta^2_A(\beta)$ reads
\beeq{
   \vartheta^2_A(\beta) \equiv \exp\Bigl[
     -\sum_{k=1}^\infty \tfrac{e^{ k\epsilon}}{k}
      \tfrac{\cosh(2k\epsilon)\cos(2\beta k\epsilon/K')-1}
           {\sinh(2k\epsilon)\cosh(k\epsilon)}
   \Bigr]\,,
}
with $\epsilon\equiv\frac{\pi K'}{K}$. 

To make sure that formula \eqref{eq:LDSF2} is correct, we have carried out  a number of checks.  First of all, one can show that the asymptotic expansion of  Eq. \eqref{eq:LDSF2} close to the Ising antiferromagnetic point are consistent with the  perturbation theory results provided in \cite{ishimura1980}. Similarly, one can perform the isotropic limit $\Delta\to -1$ to recover the previously known result \cite{Karbach1997} for the isotropic case. Alternatively, one can use the results of \cite{Caux2012} for the massless regime and take the isotropic limit.

Secondly,  to further assess the correctness of our formula we have  analysed several well-known sum rules for dynamical spin-spin correlation functions \cite{Hohenberg1974,Gerhard1982}. For simplicity, we will solely focus on  the total integrated intensity and the  first frequency moment sum rules, which are given by:
\begin{eqnarray}
\hspace{-.2in}
   a(\Delta) &\equiv&
     \int\limits_0^{\infty} \tfrac{d\omega}{2\pi}
     \int\limits_0^{2\pi} \tfrac{d Q}{2\pi}\ S^{zz}(Q,\omega)
   = \tfrac{1}{4}\,,
\label{eq:sumrule:tot} \\
\hspace{-.2in}
   g(Q,\Delta) &\equiv& 
     \int\limits_0^{\infty} \tfrac{d\omega}{2\pi}\ \omega S^{zz}(Q,\omega)
   =-2J F_{x}(1+\cos Q)\,, 
\label{eq:sumrule:E}
\end{eqnarray}
respectively. Here $F_{x}=\bracket{S^{x}_{\ell}S^{x}_{\ell+1}}$ is the nearest neighbor static correlation function for which an exact formula is known \cite{takahashi2004}.  Let us denote as $a_{(m)}(\Delta)$ and $g_{(m)}(Q,\Delta)$ the $m$-spinon contribution to each of the above two sum rules. 

\begin{figure}[t]
\begin{center}
\includegraphics[width=8cm,height=6cm]{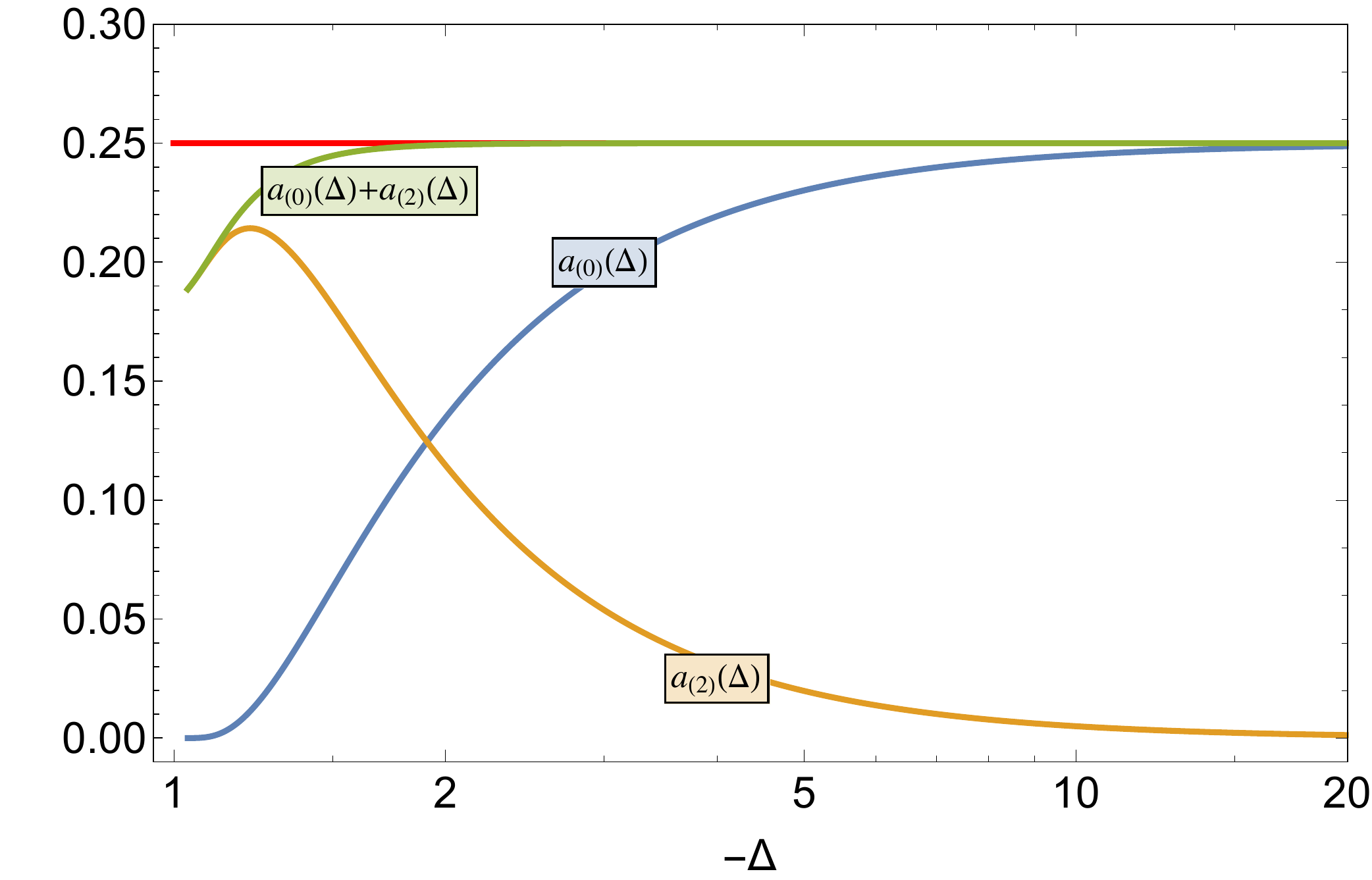}
\end{center}
\caption{
   The total integrated DSF intensity [cf. \Eq{eq:sumrule:tot}]  obtained from the exact  formulas for $a_{(m)}(\Delta)$ for $m=0,2$  (solid lines).    The solid blue line  corresponds to  the  $a_{(0)}(\Delta)$  contribution that results from the vacuum-to-vacuum transitions. The solid  orange line  represents  $a_{(2)}(\Delta)$ , while the solid green line is the sum of both contributions.  The latter  is precisely equal to 1/4 (solid  red line) for values $\Delta\lesssim -2$,  demonstrating consistency with this sum rule.}
\label{fig:2}
\end{figure}

Let us start discussing  the total integrated intensity sum rule $a(\Delta)$. This comparison is shown in Fig.  \ref{fig:2},  where we plot the various contributions of the total integrated intensity as a function of $-\Delta$.  More precisely, the blue solid line  corresponds to the zeroth contribution $a_{(0)}(\Delta)$,  which is Baxter's formula for the contribution to the staggered magnetization .  Similarly,  the solid orange line shows  $a_{(2)}(\Delta)$ coming from the theory, while the solid green line represents the sum of both contributions. We can conclude that the two-spinon contribution  carries most of the weight of the DSF  for values of $\Delta\lesssim -2$, as they saturated to the value 1/4 (shown in the figure by a solid red line) of this sum rule, while when approaching the isotropic case for $-2\lesssim \Delta\leq -1$  higher spinon excitations start contributing. This comparison with the sum rule demonstrates the correctness of our formula.

\begin{figure}[t]
\begin{center}
\includegraphics[width=8cm,height=6cm]{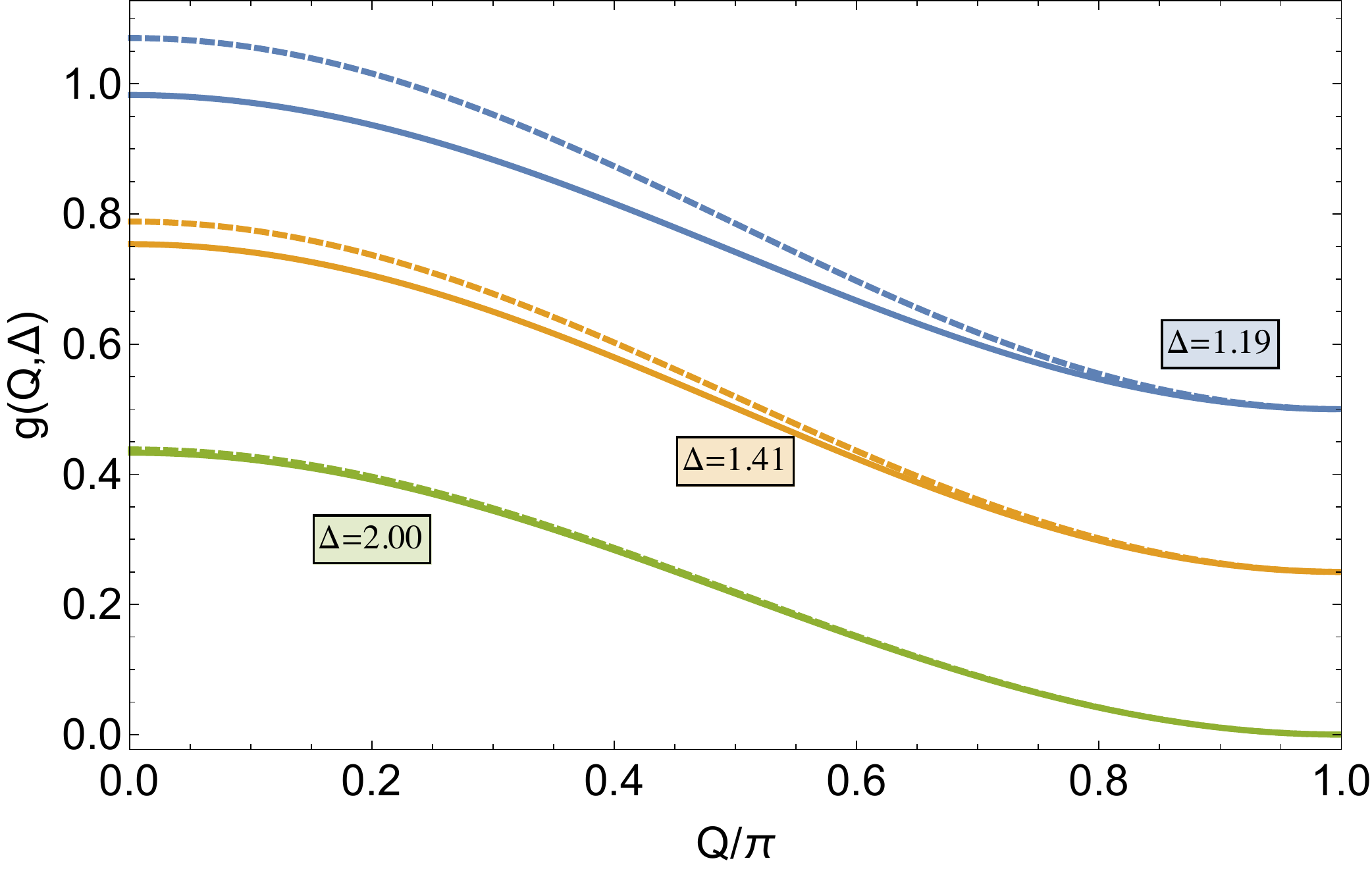}
\caption{Comparison between theory  for the first frequency moment sum rule [cf. \Eq{eq:sumrule:E}]. In all cases, the dashed  lines correspond  to the exact expression for this sum rule for $\Delta=1.19$, $1.41$, and $2.00$.   The color-matched solid lines  correspond to   $g_{(2)}(Q,\Delta)$ using the exact two-spinon contribution formula.  The curves for  $\Delta=1.19$ and $\Delta=1.41$  have been shifted upward by $1/2$ and $1/4$, respectively, so they  can be clearly discerned.}
\label{fig:3}
\end{center}
\end{figure}

Next, we compare  the first frequency moment sum rule $g(Q,\Delta)$ between theory and  the exact formula  in \Fig{fig:3} as a function of the total momentum $Q$ and for three values of the anisotropy parameter $\Delta$. The dashed line shows  the exact complete result for  $g(Q,\Delta)$ [\Eq{eq:sumrule:E}], while the solid lines with matching colors correspond to the analytical result for  $g_{(2)}(Q,\Delta)$. Similar to \Fig{fig:2}, they agree fairly well for $\Delta\lesssim -2$ where the zeroth and two-spinon contribution dominates the DSF. Note, in particular, that the value of $g_{(2)}(Q,\Delta)$ tends to zero as we approach the Ising antiferromagnetic point. This is  expected since, at this point, there is no dynamics associated to the $z$-component of the spin operator.

In this Letter, we have derived an exact and compact expression for the two-spinon contribution to the longitudinal DSF, overcoming previously unsurmountable mathematical difficulties that had hindered the study of the rich dynamical properties of XXZ model. We have made sure that our analytical findings are correct by thoroughly comparing them with  various sum rules and limiting cases.  It took around 80 years since Heisenberg's introduction of the model, to obtain a correct formula for the two-spinon contribution to the transverse DSF \cite{Perez2008} and an extra 12 years to obtain a similar formula for its longitudinal counterpart. Hopefully, with the methods developed here  more rapid advancement can be achieved on other interesting observables in order to arrive to a more complete description of this fascinating model.

\begin{acknowledgments}
IPC thanks hospitality of Brookhaven National Laboratory and acknowledges discussions, interests and help with the manuscript to Andreas Weichselbaum, Igor Zalykniak and J.S Caux. He also acknowledges partial  financial support from funding UNAM-DGAPA-PAPIIT-IN106219.
\end{acknowledgments}

\bibliography{biblio}

\end{document}